\documentstyle[11pt,pasp,twoside,epsf]{article}
\def\edcomment#1{\iffalse\marginpar{\raggedright\sl#1\/}\else\relax\fi}
\reversemarginpar

\begin{document}
\title{Cluster X-ray Luminosity Evolution}
\author{J. Patrick Henry}
\affil{Institute for Astronomy, University of Hawaii, 2680 Woodlawn Drive,
Honolulu, HI 96822, USA}

\begin{abstract}
Whether the X-ray luminosities of clusters of galaxies evolve has been
a contentious issue for over ten years. However, the data available to
addresss this issue have improved dramatically as cluster surveys from
the ROSAT archive near completion. There are now three samples of
nearby clusters and seven distant cluster samples. We present a
uniform analysis of four of the distant cluster samples. Each exhibits
highly statistically significant luminosity evolution. We combine
three of these samples to measure the high redshift cluster X-ray
luminosity function with good statistics that shows the nature of the
evolution.
\end{abstract}

\section{Introduction}

Evidence for evolution of the luminosities of clusters of galaxies
came originally from The Einstein Extended Medium Sensitivity Survey
(EMSS) (Gioia et al. 1990; Henry et al. 1992). These studies found
that the co-moving number density of high luminosity clusters is
smaller in the past than at present. Although the EMSS was the first
X-ray survey capable of finding clusters at high redshifts (here
defined to be $> 0.3$), hence the first able to search for evolution,
it did have some limitations. Perhaps the most severe was the
relatively small size of the statistical sample, 67 objects. This
limitation was compounded by the soft energy band of the EMSS, 0.3 -
3.5 keV caused by the use of focusing optics. Consequently there was
virtually no overlap with almost all previous work in X-ray astronomy
(mostly in the 2 - 10 keV band) that might have been used to augment
the sample. Thus all evidence for evolution, in particular the
comparison of low and high redshift X-ray luminosity functions (XLFs),
had to come from within the EMSS itself.  Since there were only about
20 objects each in the low and high z bins, the statistical
significance of the result was only $3 \sigma$. Consequently, the
measurement of evolution did not enjoy universal acceptance.

The advent of the ROSAT All-Sky Survey (RASS) and Pointed Program has
provided a new opportunity to study cluster luminosity evolution. The
RASS provided the huge solid angle required to construct low z samples
containing hundreds of objects. The local XLF is now determined very
reliably, with good agreement among three samples (Ebeling et
al. 1997, BCS; De Grandi et al. 1999, RASS1 BS; B\"ohringer et
 al. 2001, REFLEX).  The RASS is also being used to find very luminous
($>10^{45}$ erg s$^{-1}$) clusters, which becasue they are so rare also
requires extremely large solid angles in order to find substantial
numbers of them (Ebeling, Edge, \& Henry 2001, MACS). The ROSAT
Pointed Program enables the construction of EMSS - like surveys. There
are at least four such surveys specifically tailored to finding
distant clusters (the latest references are Nichol et al. 1999, Bright
SHARC; Vikhlinin et al. 2000, 160 deg$^2$; Rosati et al. 2000, RDCS;
Jones et al. 2000, WARPS). Finally, we are using the RASS data around
the North Ecliptic Pole to construct a survey that is both as deep as
the pointed surveys and is also contiguous (see Henry et al. 2001 for 
an overview, NEP).

\section{Uniform Analysis of Four High Redshift Cluster Samples}

Each of the surveys mentioned in the introduction has a unique
selection function that must be removed in order to compare them. The
usual method, ploting luminosity functions, does not use all the
information available. Since the evolution seems to be a lack of
objects at high redshifts, there is nothing to plot if the objects are
not there. Instead we perform maximum likelihood fits of four high
redshift samples to the AB model introduced by Rosati et al (2000).
In this model the XLF is $n(L,z) = n_{0}(z) L^{-\alpha}
e^{-L/L^{*}(z)}$, with $n_{0}(z) = n_{0} [(1+z)/(1+z_{0})]^{A}$ and
$L^{*}(z) = L^{*}_{0} [(1+z)/(1+z_{0})]^{B}$. Note that $n_{0},
\alpha,$ $L^{*}_{0},$ and $z_{0}$ are not fit, but come from a low redshift
XLF, in this case the BCS since it has the lowest normalization of the
three determinations thus yielding the least evolution. No evolution
in this model is the point $A = B = 0$. Note further that a maximum
likelihood fit incorporates the information provided by any
``missing'' high redshift clusters. We assume that $H_{0} = 50$ km
s$^{-1}$ Mpc$^{-1}$ and $q_{0} = 0.5$ to be consistent with previous
work.

\begin{figure}
\plottwo{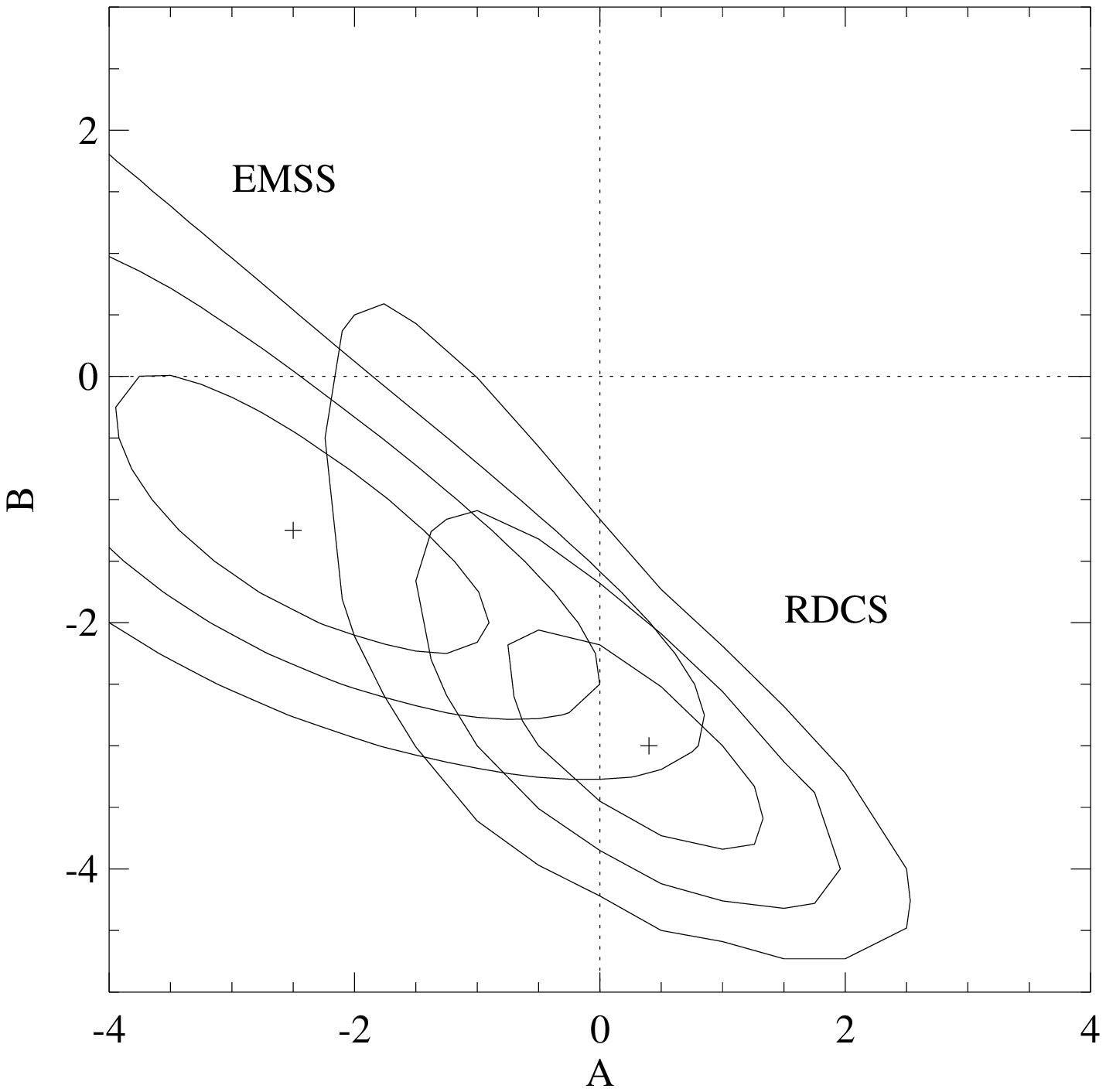}{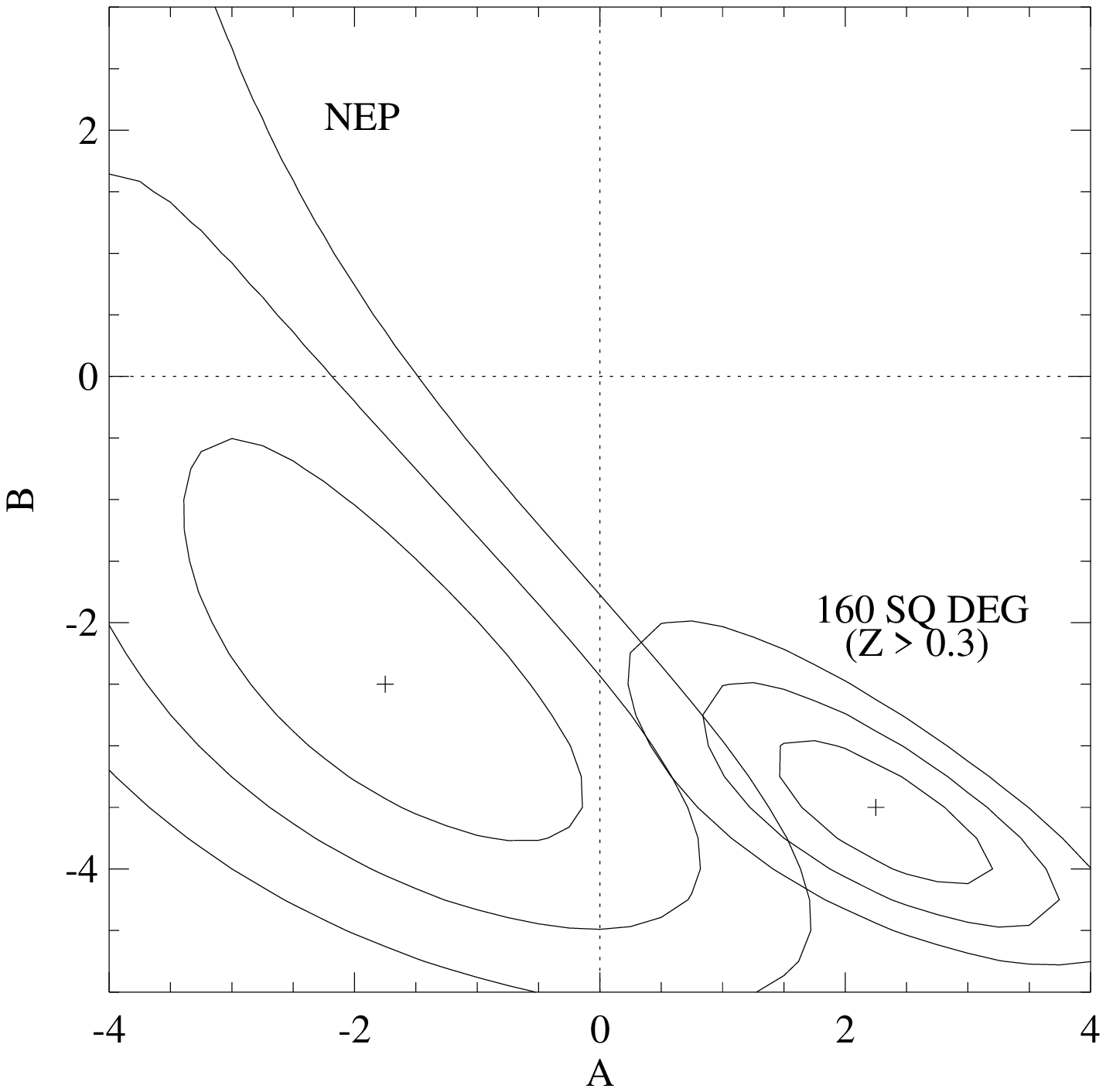}
\caption{(left) One, two and three $\sigma$ contours of the AB
model fit to the EMSS and RDCS surveys. Figure 2. (right) One, two and 
three $\sigma$ contours of the AB model fit to the 160 deg$^{2} z > 0.3$ 
and NEP surveys.}
\end{figure}

We show the results of the fits in Figures 1, 2, and 3. All four
samples exhibit luminosity evolution at the $>> 3 \sigma$
level. Figure 3 shows that the agreement among three of the surveys is
approximately at the $1 \sigma$ level but that the agreement with the
fourth is marginal. More work will be required to determine whether
this diagreement is real or results from the specific model fitted. In
particular, we have forced the best fitting low redshift XLF onto the
fit without considering the errors in its parameters.

\section{High Redshift Cluster Luminosity Function from Three Samples}
The fits described in Section 2 show that cluster luminosity evolution
is occuring. We construct the high z XLF from the sum of the EMSS,
NEP, and 160 deg$^{2}$ samples in order to obtain a higher statistics
non parametric description of that evolution. The overlap on the sky
of these three samples is about 5\%, so we have corrected
statistically for double counting since the corrections are not
large. We compare this high z XLF to the three low z XLFs in Figure
4. The high z XLF, which has a median redshift of 0.43, falls a factor
of two below the average of the three low z XLFs at a luminosity of $2
\times 10^{44}$ erg s$^{-1}$ in the 0.5 - 2.0 keV band. Further the AB
model fit to the RDCS data alone also provides a reasonable
description to the XLF of the combined sample. We emphasize that the
RDCS model is a prediction determined independently and has no
adjustable parameters.

\begin{figure}
\plottwo{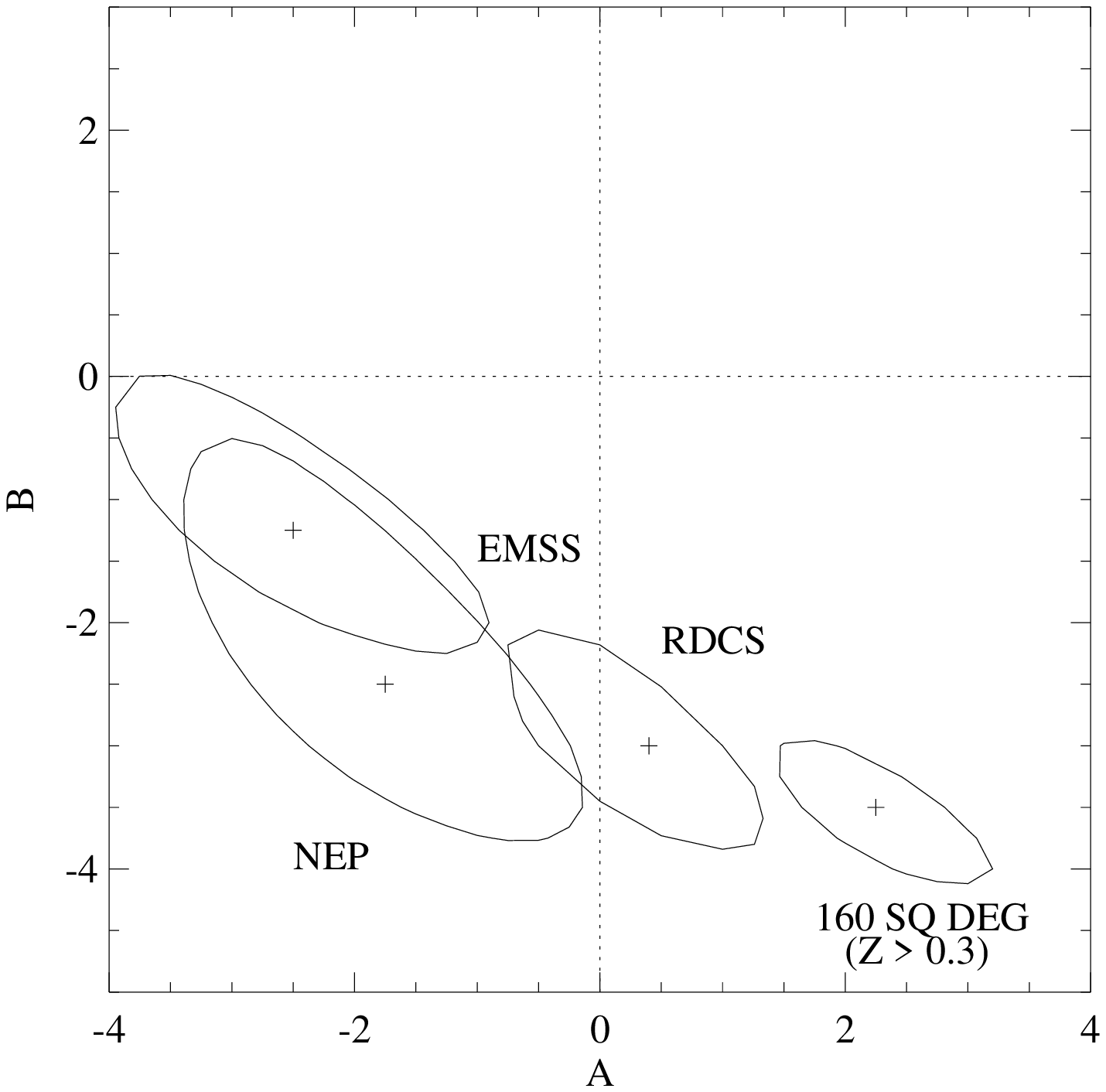}{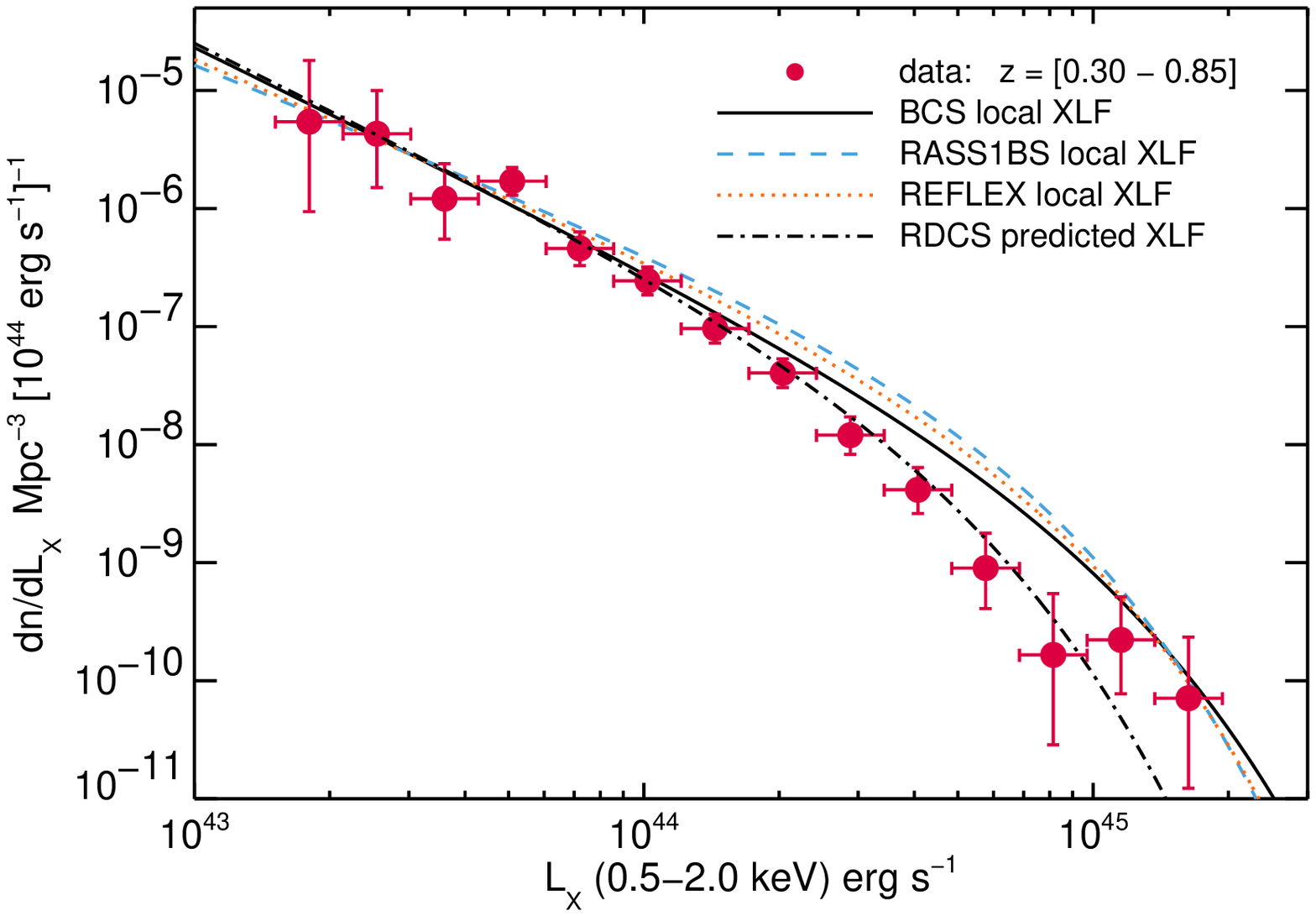}
\caption{(left) One $\sigma$ contours of the AB model fit to 
the EMSS, RDCS, NEP and 160 deg$^{2} z > 0.3$ samples. The first three 
surveys agree at this level when fit to this specific model.
Figure 4. (right) High redshift cluster luminosity function determined 
from the combined EMSS, NEP and 160 deg$^{2}$ surveys compared to three 
local luminosity functions. The RDCS best fit AB model predicts the high
redshift function well with no adjustable parameters.}
\end{figure}

\section{Conclusions}
A preliminary analysis of the MACS bright sample shows that this sample
also exhibits luminosity evolution at $> 3\sigma$. Thus there are now
five nearly independent samples of high z clusters that find evolution:
EMSS, 160 deg$^{2}$, RDCS, NEP, and MACS. We feel that the salient question
is no longer ``does cluster evolution exist?'', but rather ``what is its
amplitude as a function of redshift and luminosity?''.

\acknowledgements I want to thank my many collaborators with whom I
have worked on cluster surveys over the years. These
include: I. Gioia, C. Mullis, W. Voges, U. Briel, H. B\"ohringer and
J. Huchra for the NEP; A. Vikhlinin, C. Mullis, I. Gioia, B. McNamara,
A. Hornstrup, H. Quintana, K. Whitman, W. Forman, and C. Jones for the
160 deg$^{2}$; H. Ebeling and A. Edge for the MACS. Most of the work
described here will eventually appear as publications coauthored with
them. I thank C. Mullis for generating Figure 4.

\end{document}